\documentclass{aa}  

\usepackage{graphicx,epsfig,longtable,color}
\usepackage{txfonts}
\usepackage{xurl}
\usepackage[english]{babel}
\usepackage[colorlinks=true,     linkcolor=blue, citecolor=blue, filecolor=blue, urlcolor=blue]{hyperref}
\newcommand{\myemail}{h.stiele@fz-juelich.de}
\def\deg{\hbox{$^\circ$}}
\newcommand{\mr}{\mathrm}
\newcommand{\nh}{\hbox{$N_{\mr H}$}}
\newcommand{\hcm}[1]{$\times 10^{#1}$ cm$^{-2}$}
\def\subsun{\mbox{$_{\odot}$}}

\def\eg{e.\,g.}                                      

\def\swift{\textit{Swift}}

\def\nice{\textit{NICER}}
\def\maxi{MAXI~J1543-564}
\def\gx339{GX\,339-4}
\def\h1743{H\,1743-322}

\def\maxi15{MAXI\,J1535--571}
\def\maxi18{MAXI\,J1820+070}
\def\maxi16{MAXI\,J1631--479}
\def\sw17{Swift\,J1727.7--1613}
\def\subsun{\mbox{$_{\odot}$}}
        
\begin{document}

   \title{\nice\ and \swift/XRT monitoring of the 2023 outburst of \sw17}

   \subtitle{}

   \author{H. Stiele
          \inst{1}
          \and
          A. K. H. Kong\inst{2}
          }

   \institute{J\"ulich Supercomputing Centre, Forschungszentrum J\"ulich, Wilhelm-Johnen-Stra\ss e, 52428 J\"ulich, Germany\\
              \email{\myemail}
         \and
             Institute of Astronomy, National Tsing Hua University, No.~101 Sect.~2 Kuang-Fu Road,  30013, Hsinchu, Taiwan\\
             \email{akong@gapp.nthu.edu.tw}
             }

   \date{Received  , 2024; accepted 2024}

 \titlerunning{2023 outburst of \sw17}
 \authorrunning{Stiele, Kong}

 \abstract
   { }
   {The X-ray transient, \sw17\ was first detected on 24 August 2023 by \swift/BAT and \textit{INTEGRAL}. We investigated data from the Neutron star Interior Composition Explorer (\nice) and the Neil Gehrels \swift\ Observatory taken between August and October 2023.  }
   { We studied diagnostic diagrams, energy spectra, and short term variability. The observations cover the initial rise of the outburst in the hard state and the transition to the soft state. We focused on the evolution of quasi-periodic oscillations (QPOs) using power-density spectra and on the evolution of the spectral parameters. }
   {The overall evolution of \sw17\ is consistent with this source being a low-mass black hole X-ray binary.  Based on the Lense-Thirring precession interpretation of type-C QPOs we obtained outer radii for the hot inner flow and found that the overall evolution of these radii agrees well with the evolution of the inner disc radii obtained from fits to the energy spectra. This result holds on all times scales tested in this study and supports the Lense-Thirring precession interpretation of type-C QPOs.}
   {}


   \maketitle



\section{Introduction}
\label{Sec:Intro}
Most low-mass black hole X-ray binaries (LMXBs) belong to the class of transient sources. During their outbursts, which typically last from weeks to months, the majority of them evolve through different states \citep{2006csxs.book..157M,2010LNP...794...53B}. The outburst evolution can be studied making use of hardness intensity diagram \citep[HID;][]{2001ApJS..132..377H,2005A&A...440..207B,2005Ap&SS.300..107H,2006MNRAS.370..837G,2006csxs.book..157M,2009MNRAS.396.1370F,2010LNP...794...53B,2011BASI...39..409B}, hardness root-mean square (rms) diagram \citep[HRD;][]{2005A&A...440..207B} and rms intensity diagram \citep[RID;][]{2011MNRAS.410..679M}. Outbursts begin and end in the low-hard state (LHS) and during many outbursts a transition to the high-soft state (HSS) can be observed. During this transition two further states, the hard and soft intermediate state (HIMS and SIMS), can be observed. The LHS shows an rms of several tens of per cent and the emission is dominated by thermal Comptonisation in a hot, geometrically thick, optically thin plasma located in the vicinity of the black hole, where softer seed photons coming from an accretion disc are up-Comptonised \citep[see][for reviews]{2007A&ARv..15....1D,2010LNP...794...17G}. In the intermediate states the mass accretion rate increases. The stronger contribution of emission from the accretion disc leads to a softer hardness ratio and that is why the intermediate states are located to the left of the LHS in the HID. Regarding their timing properties the HIMS shows a reduced strength of the band-limited noise, while in the SIMS the band-limited noise is replaced by a power-law noise \citep{2011BASI...39..409B}. The HSS has an even lower variablility \citep[fractional rms $\sim$1 per cent, e.g.][]{2005A&A...440..207B} and the energy spectrum is clearly dominated by an optically thick, geometrically thin accretion disc \citep{1973A&A....24..337S}.

The different states and transitions between them are attributed to major changes in the properties of the inner accretion flow \citep[\eg\ ][]{2007A&ARv..15....1D} and they can be explained by a geometrically thin, optically thick accretion disc with an inner truncation radius varying as a function of the accretion rate \citep[\eg\ ][]{1997ApJ...489..865E,2001ApJ...555..483E}. Observations show that during the HSS the accretion disc extends down to the innermost stable circular orbit \citep[ISCO; see \eg\ ][]{2004MNRAS.347..885G,2010ApJ...718L.117S,2012PASJ...64...13N,2014SSRv..183..295M,2018arXiv181007041N,2019SCPMA..6229504D,2020MNRAS.493.5389F}. For the LHS the situation is much less clear, as different observational studies give different estimates of how close the disc comes to the black hole in this state \citep[\eg\ ][]{2006ApJ...653..525M,2014A&A...564A..37P,2010MNRAS.407.2287D,2014MNRAS.437..316K,2015A&A...573A.120P}. In the truncated disc model it is assumed that the disc recedes in the LHS and that the inner parts are filled by a radiatively inefficient, optically thin, advection-dominated accretion flow \citep[\eg\ ][]{1995ApJ...452..710N,1997ApJ...489..865E}.

The different states also show different features in power density spectra \citep[PDS;][and references therein]{2014SSRv..183...43B}. PDS of many outbursts show type-C quasi-periodic oscillations (QPOs) \citep[][and references therein]{1999ApJ...514..939W,2011MNRAS.418.2292M} in the LHS and HIMS \citep{2006csxs.book..157M,2010LNP...794...53B}. These PDS can be well fitted with a combination of Lorentzian components \citep{1994A&A...283..469V,1994ApJS...92..511V,2000MNRAS.318..361N,2002ApJ...572..392B,2005A&A...440..207B,2018ApJ...868...71S,2020ApJ...889..142S,2021ApJ...914...93S}. The QPOs are fitted with a Lorentzian from which the centroid frequency ($\nu_0$) and half width at half maximum ($\Delta$) of these oscillations can be obtained. The centroid frequencies of type-C QPOs range from 0.01 to 30 Hz, and their quality factor ($Q=\nu_0/(2\Delta)$) is $\ga10$ \citep[see \eg\ ][]{2005ApJ...629..403C,2010ApJ...714.1065R}. Often one or two overtones and at times a sub-harmonic can be observed, which are fitted with additional Lorentzians. The PDS always show band limited noise (BLN; fitted with one or more zero-centred Lorentzians) and the characteristic frequency of the QPO is anti-correlated with the total broad-band fractional rms variability. Lense-Thirring precession of a radially extended region of the hot inner flow can explain the cause of these oscillations \citep{1998ApJ...492L..59S,2009MNRAS.397L.101I}. 
The SIMS is defined by the presence of another type of QPO, the type-B QPO. Its centroid frequency, which lies between 0.8--6.4 Hz is correlated with the hard X-ray flux \citep[see ][]{2011MNRAS.418.2292M}. The Q factor is bigger than six and often an overtone and a sub-harmonic can be seen. In this state the PDS shows a power-law noise instead of the band-limited noise seen in the harder states. 
At softer states a third type of QPO can be found. These type-A QPOs are broad ($Q\sim1-3$) and weak (fractional rms < 5\%) and have centroid frequencies of 6.5--8 Hz \citep{2005ApJ...629..403C,2011MNRAS.418.2292M}.

\sw17\ was first detected by both \swift/BAT and \textit{INTEGRAL} on August 24, 2023 \citep{2023GCN.34537....1P}. The outburst of this new transient was followed by many optical \citep{2023ATel16209....1W,2023ATel16225....1B} and X-ray telescopes \citep{2023ATel16205....1N,2023ATel16207....1O,2023ATel16210....1L,2023ATel16217....1S,2023ATel16242....1D}. The detection of bright emission lines of hydrogen and helium in optical spectra led to the classification of \sw17\ as LMXB candidate  \citep{2023ATel16208....1C}. Energy spectra and temporal properties observed in the X-rays further support this classification  \citep{2023ATel16215....1P,2023ATel16219....1D}.  A study of spectral and timing properties observed during the initial outburst phase at high energies with \textit{INTEGRAL} and \textit{SRG}/ART-XC has been presented by \citet{2024MNRAS.531.4893M}. Further studies focus on investigating the X-ray polarisation during state transition \citep{2024ApJ...968...76I} and optical spectra \citep{2024A&A...682L...1M}.
In this paper, we present a comprehensive study of the temporal variability properties of \sw17\ observed during its 2023 outburst based on \nice\  and \swift/XRT data. We also include a study of the spectral properties based on \nice\ data.

\section[]{Observation and data analysis}
\label{Sec:obs}
\subsection{\nice}
The Neutron star Interior Composition Explorer \citep[\nice;][]{2012SPIE.8443E..13G} observed \sw17\ between August 25 and October 9 2023.  We used HEAsoft (v.~6.32) tasks to derive power density spectra from the pre-processed event files provided by the \nice\ datacenter. To study the low-frequency QPOs, which can appear at frequencies between $\sim0.01$ Hz and 30 Hz (see Sect.\,\ref{Sec:Intro}), we used time bins of 0.01~s and intervals of 16\,384 bins, thus covering frequencies between $6.1\times10^{-3}$ and 50 Hz. Giving the brightness of the source contribution by background photons to the PDS can be neglected. We then normalised the PDS that cover the full energy range according to \citet{1983ApJ...272..256L} and subtracted the contribution due to Poissonian noise, which has an expected value of 2 (see Sect.\ \ref{Sec:obs_sw} for a way to take instrumental effects that can cause (slight) deviations from this value into account). The PDS is then converted to square fractional rms by dividing them by the square of the intensity, which leads to the intrinsic power being independent of the rate \citep{1989ASIC..262...27V,1990A&A...227L..33B}.\footnote{\url{http://www.brera.inaf.it/utenti/belloni/ASTROSAT/Home_files/Timing.pdf}} We fitted the PDS with Lorentzians, some centred at zero and others not. The zero-centred Lorentzians were identified as BLN, while the non-zero-centred ones represented the QPOs. In 13 \nice\ observations (between September 16 and 24) (almost) no exposure is left after screening and hence we cannot include these observations in our study. 

We also studied energy spectra. They were extracted together with background and response files using the task \texttt{nicerl3-spect}. We extracted a background file based on the \textsc{scorpeon} model that can be fitted to the spectrum. The spectral fitting is done with \textsc{Xspec} \citep[V.\ 12.13.1;][]{1996ASPC..101...17A}.

\subsection{Neil Gehrels \swift\ Observatory}
\label{Sec:obs_sw}
We also analysed all \swift/XRT \citep{2005SSRv..120..165B} monitoring data of \sw17\ obtained in window timing mode between August 28 and October 24 2023. To extract energy spectra of each observation we make use of the online data analysis tools provided by the Leicester \swift\ data centre\footnote{\url{http://www.swift.ac.uk/user\_objects/}}, including single pixel events only \citep{2009MNRAS.397.1177E}.  
We used the GHATS package (v.~1.1.1), developed at INAF-OAB \footnote{\url{http://www.brera.inaf.it/utenti/belloni/GHATS\_Package/Home.html}} to obtain PDS in the 0.3 -- 10 keV energy band, following the procedure outlined in \citet{2006MNRAS.367.1113B}. The PDS, covering frequencies between $4\times10^{-3}$ Hz and 35.13 Hz, are not corrected for any contribution due to background photons. To subtract the contribution due to Poissonian noise \citep{1995ApJ...449..930Z}, we fitted the flat tail of the Leahy normalised PDS at the high-frequency end (above $\sim30$ Hz) with a constant \citep[see e.g.\ ][]{2020ApJ...889..142S,2021ApJ...914...93S}. The value of this constant gives the level of the Poissonian noise. This approach allows us to take into account deviation from the expected value of 2, that are caused by pile-up effects in the \swift/XRT data \citep{2013ApJ...766...89K}. We then converted them to square fractional rms.

\begin{figure*}
\resizebox{\hsize}{!}{\includegraphics[clip,angle=0]{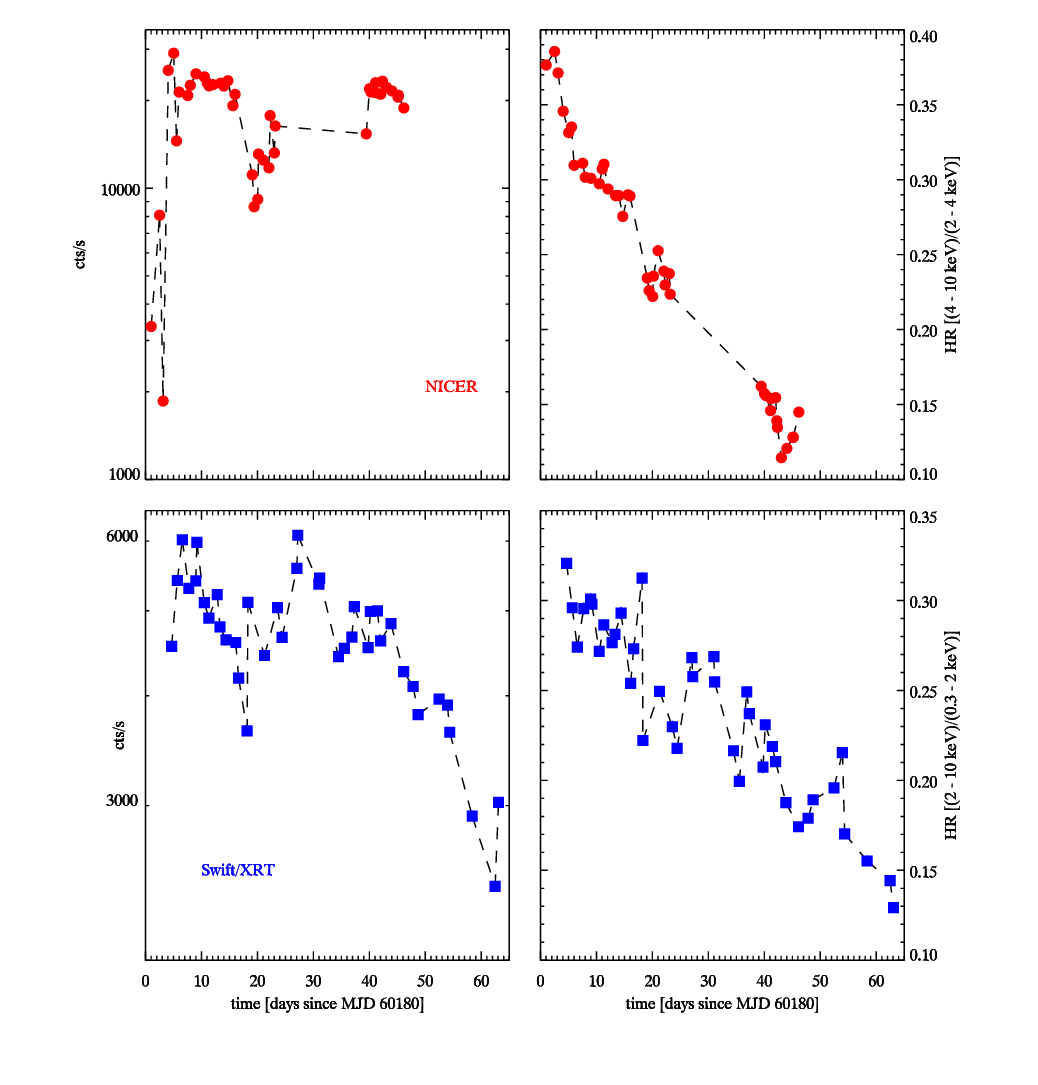}}
\caption{Light curve of the 2023 outburst of \sw17\ based on Swift/XRT and \nice\ data. Each data point represents one observation. (Green) filled triangles indicate QPOs detections. T=0 corresponds to August 24 2023 00:00:00.000 UTC.}
\label{Fig:LC}
\end{figure*}

\begin{figure*}
\resizebox{0.5\hsize}{!}{\includegraphics[clip,angle=0]{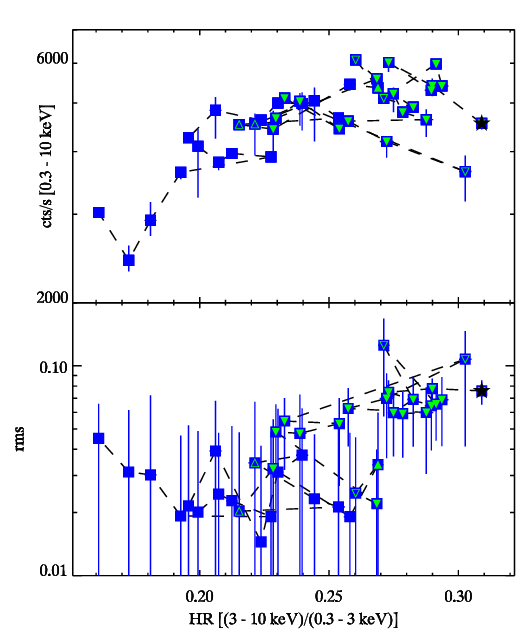}}\vspace{0.5cm}\resizebox{0.5\hsize}{!}{\includegraphics[clip,angle=0]{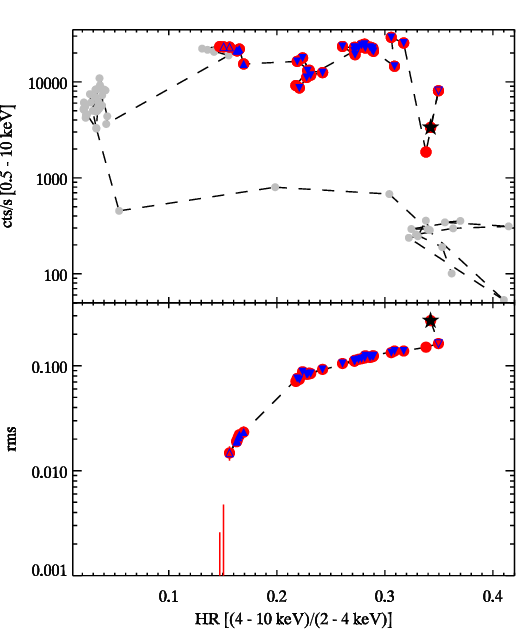}}
\caption{Hardness-intensity diagram (upper panel) and hardness-rms diagram (lower panel), derived using \swift/XRT (left panels) and \nice\ (right panels) data. Each data point represents one observation. In the HID based on \nice\ data we include further observations (grey points) to show that the overall evolution of the source shows the q-shape typically seen in HIDs of black hole LMXBs. The first observation is marked with a (black) star. Observations in which type-C QPOs are detected are marked by down-pointing triangles, while observations in which type-B QPOs are detected are marked by up-pointing triangles. The triangles are filled when the feature is detected with $\ge3\sigma$.}
\label{Fig:HID}
\end{figure*}

\begin{figure}
\resizebox{\hsize}{!}{\includegraphics[clip,angle=0]{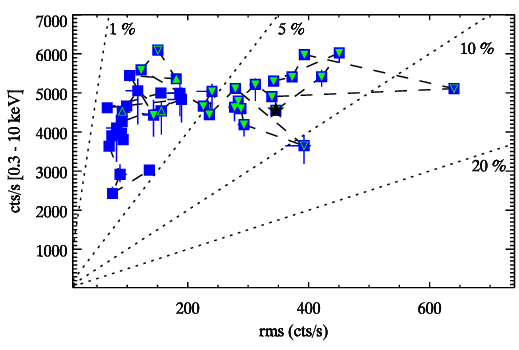}}\\ 
\smallskip
\resizebox{\hsize}{!}{\includegraphics[clip,angle=0]{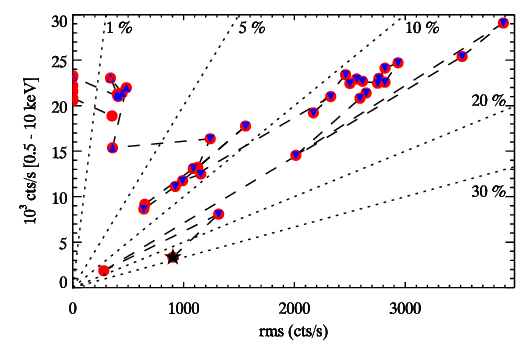}}\\
\caption{Total-rms-intensity diagram, derived using \swift/XRT (upper panel) and \nice\ (lower panel) data. Each data point represents one observation. Symbols are the same as in Fig.~\ref{Fig:HID}.}
\label{Fig:RID}
\end{figure}

\section[]{Results}
\label{Sec:res}

\subsection{Diagnostic diagrams}
\label{Sec:diag}
We derived source count rates from the \swift/XRT data in three energy bands, making use of the online data analysis tools (see Sect.~\ref{Sec:obs_sw}): total (0.3 -- 10 keV), soft (0.3 -- 3 keV), and hard (3 -- 10 keV). We also determined \nice\ count rates from the energy spectra with \textsc{Xspec} using the total (0.5 -- 10 keV), soft (0.5 -- 2 keV), medium (2 -- 4 keV), and hard (4 -- 10 keV) bands. The \swift/XRT and \nice\ light curves are shown in Fig.~\ref{Fig:LC}. Hardness ratios (HR) are obtained for \swift/XRT by dividing the count rate observed in the hard band by the one obtained in the soft band. To get HR for \nice\ data we divide the count rate in the hard band by the one of the medium band. For \swift/XRT the fractional rms is determined in the 0.3 -- 10 keV band and in the $4\times10^{-3}$ -- 35.13 Hz frequency range, while for \nice\ we use the total energy band and the $6.1\times10^{-3}$ -- 50 Hz frequency range. 

The \nice\ light curve shows the initial outburst rise, that is followed by a plateau where the count rate stays rather constant (around about 22,300 cts/s) for about 10 days. After a gap of about three days \sw17\ is observed at a lower count rate. After another gap of about 16 days, due to observations which no exposure after filtering, the source is observed again at a count rate similar to the one during the plateau phase (21,600 cts/s).   
The \swift/XRT light curve covers the finial phase of the initial rise before \sw17\ enters into the plateau phase. The lower count rates around day 20 and the the return to count rates similar to the one seen during the plateau phase around day 40 can also been observed in the \swift/XRT light curve. Furthermore, it shows that the source made an excursion towards higher count rates around day 30. While the coverage of the outburst with \swift/XRT data is less dense, it could follow the outburst for a further 17 days, showing that overall after the initial rise the count rate fluctuated around a plateau for about  38 days before the count rates decreased.

The HID (Fig.~\ref{Fig:HID}) based on \nice\ data shows that overall the source softens during the outburst with a few excursions towards higher HRs. The overall trend towards a softening of the HR is also visible in the HID based on \swift/XRT data, however it is less obvious as due to the inclusion of softer photons the excursions towards harder HRs are more numerous and more pronounced. Including further \nice\ observations up until 7 May 2024 in the HID (shown as grey points is Fig.~\ref{Fig:HID}; and otherwise not used in our study) we find that the HID shows the q-shape typically seen in black hole X-ray binaries. The observations investigated in our study cover the initial rise and the upper branch, corresponding to the transition from the hard to the soft state.

From the HRD (Fig.~\ref{Fig:HID}) obtained from \nice\ data we see that until day 23 the fractional rms shows a slow decrease with values around 10 per cent. After day 39 the rms drops to values below 2 per cent. The overall trend in the evolution of the fractional rms seen in the HRD based on \nice\ data also shows up in the HRD based on \swift/XRT observations. However, the HRD based on \swift/XRT data is less conclusive, as the rms values obtained from \swift/XRT have bigger uncertainties.

The RID (Fig.~\ref{Fig:RID}) based on \nice\ data shows that at the beginning of the outburst (until day 6) \sw17\ follows a ``hard line'', a linear relation between total rms and count rate, seen during outburst rise of black hole X-ray binaries \citep{2011MNRAS.410..679M}. During the outburst evolution when the source transits through the different states the total and fractional rms decrease. 

\subsection{Timing properties}
\label{Sec:time}

\begin{figure}
\resizebox{\hsize}{!}{\includegraphics[clip,angle=0]{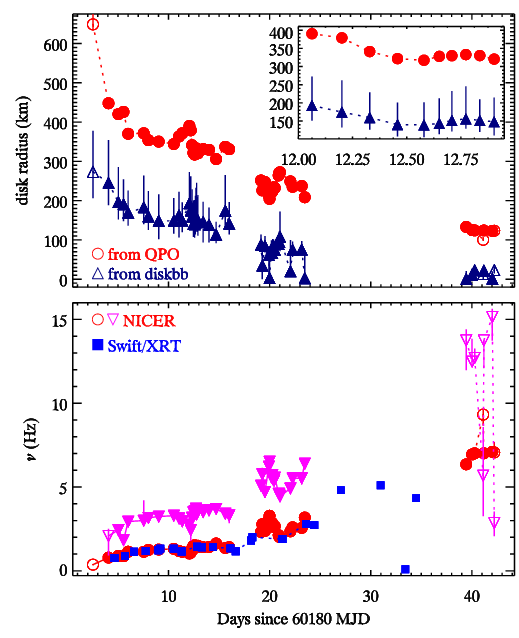}}
\caption{Lower panel: Evolution of the characteristic frequency of the QPOs during the outburst. Circles indicate fundamental QPOs from \nice\ observations, while down-pointing triangles indicate upper harmonics. Fundamental QPOs from \swift/XRT observations are indicated by squares.
Upper panel: Comparison of the radius derived from the characteristic frequency of the fundamental QPO following \citet{2009MNRAS.397L.101I} and from the normalisation of the \texttt{diskbb} model \citep{1984PASJ...36..741M} fitted to the energy spectra obtained from \nice\ observations. The inset exemplifies the evolution within one day. In both panels filled symbols indicate observations or snapshots in which the QPO is observed at $\ge3\sigma$.}
\label{Fig:nu_qpo_ni_ev}
\end{figure}

\begin{figure}
\resizebox{\hsize}{!}{\includegraphics[clip,angle=0]{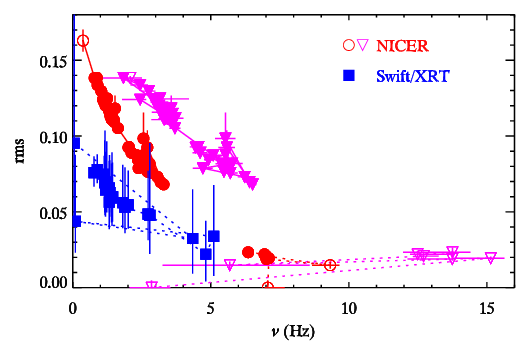}}
\caption{Correlation of the broadband fractional rms variability with the characteristic QPO frequency, derived from \nice\ and \swift/XRT data. Symbols are the same as in the lower panel of Fig.~\ref{Fig:nu_qpo_ni_ev}}
\label{Fig:fchar_rms_ni}
\end{figure}

\subsubsection{\nice}
\label{Sec:time_ni}
In the \nice\ observation taken on day 2 the PDS shows a BLN component and a QPO with a characteristic frequency of 0.37 Hz and a Q factor of 7.9 but at a low significance of 2.0 $\sigma$. The PDS of the \nice\ observations taken between day 4 and until day 42 can be well described by two BLN components and two QPOs. In the \nice\ observations taken on days 12, 19, 20, and 23, the fundamental QPO appears as a double-peak feature with a separation in the characteristic frequencies of the two peaks of less than 1 Hz. We split these observations into snapshots \citep[see][]{2023MNRAS.522..268S} and studied the PDS of the individual snapshots. Within the PDS of the individual snapshots we always find a well defined single-peak QPO. This shows that the characteristic frequency of the QPO can change by up to some tenth of Hz on timescales of a few hours.

The evolution of the characteristic frequency of the QPOs are shown in Fig.\ref{Fig:nu_qpo_ni_ev}. The characteristic frequency of the fundamental QPO increases from $\sim$0.4 Hz on day 2 to a plateau at about 1.3 Hz between days 6 and 16. During this time the characteristic frequency of the fundamental QPO fluctuates between 1.0 and 1.6 Hz. Between days 19 and 23 the characteristic frequency varies around an average value of 2.6 Hz (2.0 -- 3.3 Hz). Between days 39 and 42 values between 6.4 and 9.3 Hz are observed. The Q factor ranges between 4 and 13 and the significance of detection is always above 3$\sigma$, apart from the observations on days 2.52, 41.12 and 42.22 (see Table~\ref{Tab:qpo_ni}). The second QPO has a characteristic frequency that increases from $\sim$2 Hz on day 4 to a plateau at $\sim3.3$ Hz between days 6 and 16. In observations taken between days 19 and 23 it varies between 4.4 and 6.5 Hz. The observed Q factors are lower with values between 0.5 and 7.1 for the observation taken until day 23. Apart from the observation on day 4, this feature is also detected with a significance $>3\sigma$ in these observations. After day 39 a second QPO is detected at low significance ($<1.7\sigma$) and the characteristic frequency varies between 2.9 and 15.1 Hz (Table~\ref{Tab:qpo_ni_uh}).   

The correlation between broadband fractional rms and characteristic frequency is shown in Fig.~\ref{Fig:fchar_rms_ni}. The observed anti-correlation between rms and frequency suggests the oscillations  up until day 23 are type-C QPOs. For the oscillations observed between days 39 and 42 with characteristic frequencies between 5 and 7 Hz the total rms is about 2 per cent. The values of the characteristic frequency and the flat distribution hints at these oscillations being type-B QPOs.  In the observations taken after day 42 no QPOs are detected. 

One model to explain type-C QPOs is the Lense-Thirring precession of a hot inner flow in a truncated accretion disc. Taking the equations and parameter values ($\zeta=0$; $h/r=0.2$) given in \citet{2009MNRAS.397L.101I} we can transfer the characteristic frequency of the fundamental QPO to the outer radius of the hot inner flow. We obtain values between $44 R_g$ and $6 R_g$. The evolution of the radii is shown in the upper panel of Fig.~\ref{Fig:nu_qpo_ni_ev} (red circles). During the plateau between days 6 and 16 the radius derived from the QPO frequency fluctuates around $23 R_g$. Superimposed are the variations on hours timescale in the QPO frequency and hence radius, shown as an example for day 12 in the inset of Fig.~\ref{Fig:nu_qpo_ni_ev}.
   
As there is an overall anti-correlation between the frequency and the radius, changes at lower frequencies imply larger changes in the radius. Depending on the spin parameter and the mass of the black hole the change in characteristic frequency from 1.09 to 1.34 Hz within 3.08 hours (on day 12) corresponds to a change in the radius of up to $2.66 R_g$ or $12\,717$ m h$^{-1}$ (for $a=0.9$ and $M_{\mr{BH}}=10 M\subsun$). The average velocity derived from the change of the characteristic frequency from 1.14 to 1.41 Hz within the 10 days of the plateau is 165 m h$^{-1}$. If the type-C QPOs are related to the radius of the hot inner flow, there must be a mechanism that causes rather fast changes of the radius on hours timescale while the overall evolution of the radius over several days progresses much slower . 

An alternative explanation for type-C QPOs is a model based on the concept of radiative feedback between hot electrons in the corona and soft radiation from the accretion disc \citep{2022A&A...662A.118M}. The electron population is heated by some unspecified mechanism that injects a fraction $f\le1$ of the energy being released when mass is accreted from infinity to the radius of the innermost stable orbit. In this model of \citet{2022A&A...662A.118M} the corona is modelled as a sphere of radius $r_c$, while the soft radiation is assumed to be produced within a sphere of radius $r_d$. This radiation is a result of the reprocessing of hard coronal radiation. Furthermore, damped oscillations, which are identified with type-C QPOs, have been observed. As a result of the damping, the oscillations must be sustained by small fluctuations in the accretion rate. To quench the type-C QPOs, the electrons must rapidly depart from the corona. This occurs as the corona contracts and the outflow narrows progressively as the source transitions from the hard to the soft state. According to Eq. (15) in \citet{2022A&A...662A.118M}, which was obtained from a stability analysis of the energy balance of the electrons and the hard photon energy density, the natural frequency of the system $f_0\propto r_c^{-3/2}$, where $r_c$ is the radius of the corona. Using the values given in \citet{2022A&A...662A.118M} we find that the fast change in the characteristic frequency on hours timescale corresponds to a change in the radius of $87\,253$ m h$^{-1}$, while the average change within the 10 days relates to $1\,119$ m h$^{-1}$. So again the model requires a mechanism to explain the highly variable frequency and radius on hours timescale while these quantities evolve more steady over several days.

\subsubsection{\swift}
\label{Sec:time_sw}
The PDS of the \swift/XRT observations taken between day 4 and 24 can be well described by up to two BLN components and a QPO. In some of the observations an upper harmonic can be detected. The characteristic frequency of the fundamental QPO increases from 0.77 Hz on day 4 to $\sim2.8$ Hz on day 24, also showing a plateau between day 6 and 16, although it is less pronounced than in the \nice\ data as there are fewer \swift/XRT observations (Fig.\ref{Fig:nu_qpo_ni_ev}). In most of the observations the Q factor is between $\sim7-10$ and apart from two observations the significance is $>3\sigma$. The upper harmonics are observed at characteristic frequencies between $\sim1.8$ and $\sim4$ Hz, with Q factors between 1.7 and 8, and a significance between about 1.4 and 3 $\sigma$. 

The PDS of \swift/XRT observations taken between day 27 and 32 show one noise component and a QPO with characteristic frequency increasing from 4.8 and 6.6 Hz. The Q factor is $>8$ and the significance between 0.9 and $4.3\sigma$. The correlation between total fractional rms and characteristic frequency is shown in Fig.~\ref{Fig:fchar_rms_ni}. The QPOs up to and including day 27 are type-C QPOs, while the QPOs seen on days 30 to 32 are of type-B. 

In the PDS of the \swift/XRT observation taken on day 33 a QPO-like feature ($Q=6.08$) that shows a ridge towards lower frequencies at a characteristic frequency of $94\pm5$ mHz is detected at $3.78\sigma$. On the next day the PDS shows BLN and a type-C QPO ($Q=18.7$) with a characteristic frequency of $4.34\pm0.05$Hz at $3.45\sigma$. On days 35 and 39 a type-B QPO may be present in the PDS, however the significance of detection is low ($\la1.5\sigma$). The PDS of day 60 shows two QPO-like feature at characteristic frequencies of $55\pm4$ mHz and $102\pm5$ mHz, respectively. The feature at lower frequency has a Q-factor of 3.7 and is detected at $3.3\sigma$, while the other one has a Q-factor of 7.6 and is detected at $2.7\sigma$.

\subsection{Spectral properties}
\label{Sec:spec}
\subsubsection{\nice}

For the \nice\ observations energy spectra in the 0.5 -- 10 keV range are fitted with \textsc{Xspec} \citep{1996ASPC..101...17A}.\footnote{The documentation of the \textsc{scorpeon} background model, which can be found here: \url{https://heasarc.gsfc.nasa.gov/docs/nicer/analysis_threads/scorpeon-xspec/}, recommends to use a broad energy range to constrain the parameters of this background model well.} For the four observations where the characteristic frequency of the QPO varied so significantly between individual snapshots that the PDS of the entire observation displayed double peaked QPOs (see Sect.\ \ref{Sec:time_ni}), we extracted and analysed the energy spectra of individual snapshots. The spectra are grouped in an ``optimal'' way based on \citet{2016A&A...587A.151K} so that the best fit can be obtained using $\chi^2$ minimisation. Systematic uncertainties of one per cent are assumed.

The spectral background is described by the \textsc{scorpeon} model (v22)\footnote{\url{ https://heasarc.gsfc.nasa.gov/docs/nicer/analysis_threads/scorpeon-overview/}} that is fitted to the energy spectrum together with the model components describing the source emission. To model the source spectrum we used an absorbed \texttt{diskbb} model \citep{1984PASJ...36..741M} together with the Comptonisation component \texttt{nthcomp} \citep{1996MNRAS.283..193Z,1999MNRAS.309..561Z}. The seed photon temperature of the latter component is set to the inner disc temperature. The foreground absorption is modelled with \texttt{tbabs} \citep{2000ApJ...542..914W}, using the abundances of \citet{2000ApJ...542..914W} and the cross sections given in \citet{1996ApJ...465..487V}. In some observations an additional \texttt{gaussian} component is included to model residuals around 6.6 keV related to the iron line.

\begin{figure}
\centering
\resizebox{\hsize}{!}{\includegraphics[clip,angle=0]{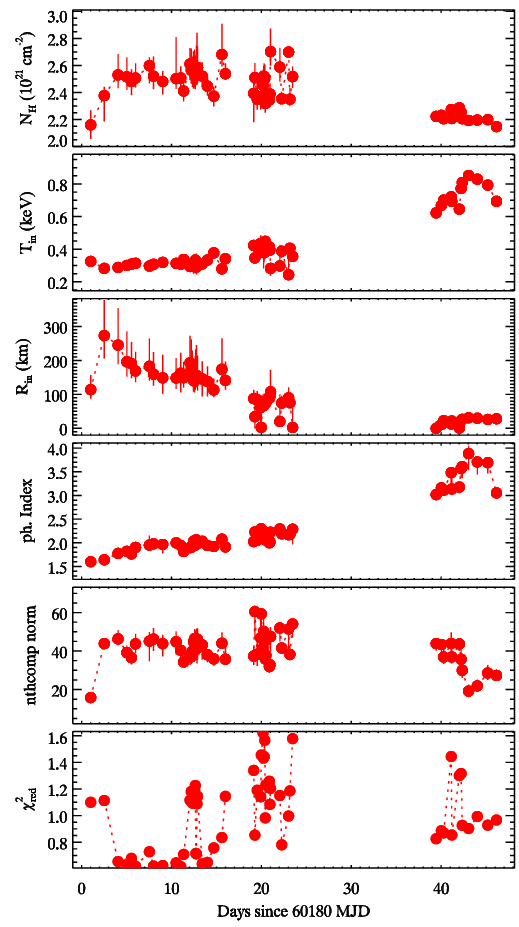}}
\caption{Evolution of spectral parameters, fitting \nice\ spectra with an absorbed disc blackbody and Comptonisation model. Given parameters are (from top to bottom): foreground absorption, inner disc temperature, inner disc radius, photon index, \texttt{nthcomp} normalisation, and reduced $\chi^2$. The electron temperature is not shown as it is not well constrained. }
\label{Fig:spec}
\end{figure}

With this model we obtain acceptable fits for all observations. Observations for which a higher reduced $\chi^2$ value is obtained show stronger residuals around the gold edge at $\sim2.2$ keV. For the observation that shows the strongest residuals, we repeated the fit excluding the 2.1--2.3 keV range and obtained spectral parameter values that are consistent within errors with the original ones. So we decided to work with the original fits (and treating all observations the same) as the outlying data points around the gold edge seem to increase only the reduced $\chi^2$ value but do not change the spectral parameter values significantly. 
The evolution of the spectral parameters during the outburst is shown in Fig.~\ref{Fig:spec}. The obtained averaged foreground absorption \nh$\sim2.43$\hcm{21} is similar to that of other Galactic X-ray binaries \citep{2015MNRAS.452.3666S}. As we modelled the spectra individually some variation in this parameter (within error bars) is to be expected \citep{2018ApJ...868...71S} and does not indicated any actual change of the foreground absorption. The temperature of the accretion disc remains rather constant at about 0.31 keV up until about day 16. Between days 19 and 23 it is more variable with an average value of $\sim0,.38$ keV. After day 39 it increases to values between 0.62 and 0.85 keV. From the normalisation of the \texttt{diskbb} model we can derive the disc inner radius using a distance of $2.7\pm0.3$ kpc \citep{2024A&A...682L...1M}. For the inclination of \sw1727\ we use $\theta=40_{-0.8}^{+1.2}$\deg reported by \citet{2024ApJ...960L..17P} and take into account a possible higher inclination of $\theta=47.9\pm0.03$\deg \citep{2023ATel16219....1D} when determining error bars. The inner radius of the disc first increases from about 114 km to a maximum of about 273 km at day 2.5 and then shows a decay to a plateau around 154 km between days 6 and 16. Between days 19 and 24 the inner disc radius appears to be rather variable around an average radius of 74 km. After day 39 the radius shows another plateau at about 23 km. The obtained photon index increases from 1.6 at the beginning of the outburst to $\sim$2.3 on day 23. After day 39 photon indices between 3.0 and 3.9 are obtained. Trying to fit the observations taken between days 39 and 46 with only an absorbed \texttt{diskbb} component results in statistically unacceptable fits with strong residuals. The normalisation of the \texttt{nthcomp} component is rather constant with values around 41. It fluctuates more between days 19 and 23 and shows an excursion towards lower values after day 43. The electron temperature is not well constrained and for many observations only lower limits can be obtained. This has to be expected as the missing coverage of higher energies above 10 keV in \nice\ spectra leads to lower values of the electron temperature.

\begin{figure}
\centering
\resizebox{\hsize}{!}{\includegraphics[clip,angle=0]{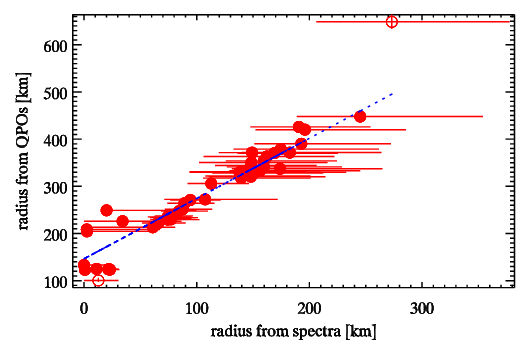}}
\caption{Plotted are outer radii of the hot inner flow obtained from the characteristic frequency of the fundamental QPO versus inner disc radii obtained from spectral fits. There is a tight correlation between these two radii for observations in which type-C QPOs are observed that holds on hours as well as days timescales. The (blue) dashed line indicates the best fitted linear correlation using only observations in which the fundamental QPO is detected with more than 3$\sigma$ (filled circles). }
\label{Fig:rad_corr}
\end{figure}

\section[]{Discussion}
\label{Sec:dis}
We investigate the \swift/XRT and \nice\ monitoring observations of the 2023 outburst of \sw17, which was first detected by \swift/XRT and \textit{INTEGRAL} on August 24, 2023. It was followed up by \swift\ and \nice\ until October 2023.

The HIDs show that \sw17\ evolves during the part of its 2023 outburst that is covered by \nice\ and \swift\ observations from the LHS through the HIMS and SIMS into the HSS. This evolution agrees with the q-shape or turtle-head pattern observed in many black hole X-ray binary outbursts. 

The RID shows that in the first six observations \sw17\ follows a ``hard line'' with a rather low fractional rms around 15\%. Typically, the ``hard-line'' is observed at higher fractional rms values of about 30 -- 40\% \citep[see \eg\ ][]{2011MNRAS.410..679M,2020ApJ...889..142S,2020MNRAS.497.3896A,2020MNRAS.499..851Z,2021ApJ...920..121Y}. During later parts of its outburst the total rms decreases together with the fractional rms.

For observations taken between day 2 and day 42, QPOs have been detected in the \nice\ data. Regarding \swift/XRT PDS, QPOs are detected with a detection significance above 3$\sigma$ between day 4 and day 34. The centroid frequency of the fundamental QPO and its evolution during the outburst agrees well in the PDS of the two instruments. 
In the \nice\ PDS a harmonic at higher frequencies can be observed. During the outburst a clear evolution of the fundamental QPO centroid frequency from $\sim0.4$ to $\sim3.2$ Hz on day 23 can be observed. The frequency range in which the QPOs are observed, the Q factors, and the anti-correlation between total fractional rms and characteristic frequency suggest these QPOs to be type-C QPOs, consistent with the detection of these QPOs during the hard state. 

The properties of the energy spectra obtained form \nice\ data confirm that \sw17\ is in the LHS or HIMS during these observations. The observed photon index increases form 1.6 to 2.3 and the disc temperature is rather low with values between 0.2 and 0.5 keV.

In the \swift/XRT observations on days 27 and 30 QPOs with a characteristic frequency $\sim5$Hz are observed. These oscillations are type-B QPOs indicating that \sw17\ is in the SIMS. On day 34 the source returned to the HIMS showing a type-C QPO with characteristic frequency of $4.34\pm0.05$ Hz. Further (type-B) QPOs with a characteristic frequency between 6.4 and 7.1 Hz are observed in \nice\ PDS between day 39 and 42. It is known that black hole LMXBs can go back and forth a few times between the HIMS and SIMS, leading to the idea that the type-C and type-B QPOs observed in these states, respectively, are caused by two distinct mechanisms \citep{2011MNRAS.418.2292M}. For a recent study on transitions in the IMSs of \gx339\ using \nice, refer to \citet{2023MNRAS.522..268S}. Therefore, the return to the HIMS on day 34 is typical behaviour for a black hole LMXB. \sw17\ then enters into the soft state, where no QPOs are observed.

Between day 6 and 16 there is little overall evolution in the characteristic frequency of the type-C QPOs with an increase of the frequency of the fundamental QPO from $\sim1.1$ to $\sim1.5$ Hz \citep[see also][]{2024MNRAS.531.4893M}. Superimposed on this plateau are faster fluctuations in the characteristic frequency, especially on day 12 where the frequency evolves in a similar range just within one day. The PDS obtained from \nice\ observations between days 19 and 23 show an overall increase in the characteristic frequency of the type-C QPO that is again superimposed by faster fluctuations of the frequency.    

Within the Lense-Thirring precession of a hot inner flow in a truncated accretion disc the frequency of the fundamental type-C QPO can be translated into the outer radius of the hot flow. Following the equations given in \citet{2009MNRAS.397L.101I} and assuming a ten solar mass black hole and a spin parameter of $a=0.9$ we derived these radii from the observed QPO frequencies. The radii obtained from the characteristic frequency of the fundamental QPO can be compared to the one derived from the normalisation of the \texttt{diskbb} model. For the later ones we use a distance of $2.7\pm0.3$ kpc \citep{2024A&A...682L...1M} and assume an inclination of $\theta=40_{-0.8}^{+7.93}$\deg \citep{2023ATel16219....1D,2024ApJ...960L..17P}.  From the upper panel of Fig.~\ref{Fig:nu_qpo_ni_ev} we see that the overall trend of the radii obtained by the two methods are similar, especially for observations up until day 16. During the plateau in the QPO frequency between days 6 and 16 and the evolution of the inner disc radius obtained from spectral fits of the \nice\ observations fluctuates around a plateau of $\sim135$ km. 
In Fig.~\ref{Fig:rad_corr} we plotted the two radii versus each other and find a tight correlation between them that holds for the overall long-term evolution as well as for the evolution on hours timescale within a day. The consistent co-evolution of the inner disc radii and the outer radii of the accretion flow supports the Lense-Thirring precession interpretation. Nevertheless there remains the challenge to explain how the rather fast changes in the QPO frequency and disc radius on hours timescale go together with the much slower and steadier evolution of these properties on timescales of days. The data points for observations after day 39 do not seem to follow the correlation, which can be addressed to the fact that the type-B QPOs that are present in these observations are thought to be caused by a different mechanism other than Lense-Thirring precession. During the SIMS when type-B QPOs exist it is assumed that the truncation radius of the disc moves inwards while the hot flow disappears. 

We refrain from comparing the numerical values of the radii determined with the two different methods, as these values depend on the assumed parameters.

The spectral properties of observations taken after day 39 confirm that \sw17\ went into softer states during these observations, as the disc temperature increased to values between 0.6 keV and 0.9 keV. For the photon index rather high values between 3 and 4 are observed. In observations in which type-B QPOs are found the source resides in the SIMS.

\begin{acknowledgements}
We acknowledge that this research makes use of SciServer, a resource developed and operated by the Johns Hopkins University, Institute for Data Intensive Engineering and Science (IDIES).
HS acknowledges support from the ``Big Bang to Big Data'' (B3D) project, the NRW cluster for data-intensive radio astronomy, funded by the state of North Rhine-Westphalia as part of the \textit{Profiling 2020} programme. AK acknowledges support by the Ministry of Science and Technology of the Republic of China (Taiwan) through grants 110-2628-M-007-005 and 111-2112-M-007-020.
This research has made use of data obtained through the High Energy Astrophysics Science Archive Research Center Online Service, provided by the NASA/Goddard Space Flight Center.
\end{acknowledgements}

%
   \bibliographystyle{aa} 
   \bibliography{} 
%

\begin{appendix} 

\begin{table*}
\caption{Parameters of the fundamental QPO in the PDS derived from \nice\ data}
\begin{center}
\begin{tabular}{rrrrrrr}
\hline\noalign{\smallskip}
 \multicolumn{1}{c}{Obs ID} &  \multicolumn{1}{c}{day$^{\dagger}$} & \multicolumn{1}{c}{$\nu_{\mr{0; QPO1}}$ (Hz)} & \multicolumn{1}{c}{$\Delta_{\mr{QPO1}}$ (Hz)} & \multicolumn{1}{c}{rms$_{\mr{QPO1}}$ (\%)} & \multicolumn{1}{c}{Q$_{\mr{QPO1}}$}  & \multicolumn{1}{c}{$\sigma_{\mr{QPO1}}$}\\ 
\hline\noalign{\smallskip}
6203980102	&2.52	&$0.369_{-0.026}^{+0.012}$ &	$0.047_{-0.031}^{+0.053}$ &	$5.63_{-1.40}^{+2.73}$ & 	$7.90_{-5.86}^{+9.23}$	&2.02	\\
\smallskip
6203980104	&4.06	&$0.780_{-0.009}^{+0.010}$ &	$0.129_{-0.025}^{+0.022}$ &	$9.55_{-0.70}^{+0.48}$ & 	$6.05_{-1.24}^{+1.10}$	&6.81	\\
\smallskip
6203980105	&5.04	&$0.888_{-0.005}^{+0.005}$ &	$0.121_{-0.012}^{+0.014}$ &	$8.86_{-0.28}^{+0.27}$ & 	$7.35_{-0.80}^{+0.89}$	&16.10	\\
\smallskip
6703010101	&5.55	&$0.865_{-0.008}^{+0.008}$ &	$0.117_{-0.014}^{+0.014}$ &	$9.20_{-0.22}^{+0.38}$ & 	$7.41_{-0.93}^{+0.98}$	&20.90	\\
\smallskip
6203980106	&6.01	&$1.128_{-0.005}^{+0.006}$ &	$0.194_{-0.013}^{+0.014}$ &	$8.78_{-0.18}^{+0.17}$ & 	$5.80_{-0.41}^{+0.43}$	&24.84\\	
\smallskip
6203980107	&7.54	&$1.125_{-0.006}^{+0.006}$ &	$0.168_{-0.013}^{+0.014}$ &	$8.84_{-0.22}^{+0.22}$ & 	$6.69_{-0.56}^{+0.59}$	&19.65\\	
\smallskip
6203980108	&8.00	&$1.234_{-0.004}^{+0.004}$ &	$0.174_{-0.010}^{+0.010}$ &	$8.56_{-0.15}^{+0.14}$ & 	$7.10_{-0.42}^{+0.44}$	&29.45	\\
\smallskip
6203980109	&9.03	&$1.256_{-0.004}^{+0.004}$ &	$0.192_{-0.008}^{+0.009}$ &	$8.56_{-0.12}^{+0.12}$ & 	$6.53_{-0.30}^{+0.31}$	&35.98\\	
\smallskip
6203980110	&10.52	&$1.305_{-0.004}^{+0.004}$ &	$0.162_{-0.008}^{+0.009}$ &	$8.27_{-0.14}^{+0.14}$ & 	$8.07_{-0.44}^{+0.47}$	&28.71\\	
\smallskip
6203980111	&11.04	&$1.173_{-0.004}^{+0.004}$ &	$0.172_{-0.008}^{+0.009}$ &	$8.36_{-0.14}^{+0.14}$ & 	$6.80_{-0.35}^{+0.37}$	&30.16	\\
\smallskip
6703010103	&11.35	&$1.127_{-0.006}^{+0.006}$ &	$0.150_{-0.016}^{+0.017}$ &	$8.38_{-0.26}^{+0.26}$ & 	$7.53_{-0.84}^{+0.91}$	&16.07\\	
\smallskip
6203980112	&12.06	&$1.029_{-0.009}^{+0.010}$ &	$0.107_{-0.020}^{+0.023}$ &	$8.38_{-0.52}^{+0.49}$ & 	$9.59_{-1.84}^{+2.14}$	&8.10\\	
\smallskip
6203980112	&12.20	&$1.091_{-0.020}^{+0.022}$ &	$0.085_{-0.037}^{+0.037}$ &	$8.05_{-1.06}^{+0.92}$ & 	$12.89_{-5.83}^{+5.86}$	&3.80	\\
\smallskip
6203980112	&12.33	&$1.324_{-0.024}^{+0.024}$ &	$0.171_{-0.030}^{+0.036}$ &	$7.17_{-0.55}^{+0.50}$ & 	$7.75_{-1.51}^{+1.75}$	&6.57\\	
\smallskip
6203980112	&12.46	&$1.481_{-0.013}^{+0.013}$ &	$0.158_{-0.026}^{+0.032}$ &	$7.85_{-0.47}^{+0.45}$ & 	$9.36_{-1.61}^{+2.00}$	&8.32\\	
\smallskip
6203980112	&12.58	&$1.513_{-0.015}^{+0.016}$ &	$0.211_{-0.044}^{+0.060}$ &	$7.96_{-0.58}^{+0.56}$ & 	$7.16_{-1.57}^{+2.11}$	&6.86	\\
\smallskip
6203980112	&12.65	&$1.426_{-0.015}^{+0.016}$ &	$0.177_{-0.032}^{+0.036}$ &	$7.88_{-0.46}^{+0.43}$ & 	$8.04_{-1.55}^{+1.73}$	&8.55\\	
\smallskip
6203980112	&12.71	&$1.413_{-0.013}^{+0.013}$ &	$0.167_{-0.030}^{+0.037}$ &	$7.89_{-0.51}^{+0.48}$ & 	$8.47_{-1.61}^{+1.98}$	&7.81\\	
\smallskip
6203980112	&12.77	&$1.390_{-0.013}^{+0.013}$ &	$0.156_{-0.030}^{+0.037}$ &	$7.70_{-0.50}^{+0.47}$ & 	$8.91_{-1.81}^{+2.19}$	&7.71	\\
\smallskip
6203980112	&12.84	&$1.411_{-0.010}^{+0.010}$ &	$0.164_{-0.022}^{+0.024}$ &	$7.87_{-0.34}^{+0.33}$ & 	$8.62_{-1.22}^{+1.35}$	&11.69\\	
\smallskip
6203980112	&12.90	&$1.491_{-0.009}^{+0.009}$ &	$0.166_{-0.022}^{+0.027}$ &	$7.67_{-0.32}^{+0.31}$ & 	$9.00_{-1.24}^{+1.52}$	&12.09\\	
\smallskip
6203980113	&13.42	&$1.398_{-0.005}^{+0.005}$ &	$0.192_{-0.013}^{+0.014}$ &	$7.86_{-0.16}^{+0.16}$ & 	$7.27_{-0.51}^{+0.54}$	&24.75	\\
\smallskip
6203980114	&14.00	&$1.392_{-0.007}^{+0.007}$ &	$0.303_{-0.016}^{+0.017}$ &	$7.96_{-0.14}^{+0.14}$ & 	$4.59_{-0.27}^{+0.28}$	&27.56\\	
\smallskip
6703010104	&14.71	&$1.624_{-0.010}^{+0.010}$ &	$0.188_{-0.025}^{+0.030}$ &	$7.39_{-0.33}^{+0.31}$ & 	$8.62_{-1.19}^{+1.44}$	&11.33\\	
\smallskip
6203980115	&15.61	&$1.348_{-0.014}^{+0.014}$ &	$0.219_{-0.029}^{+0.035}$ &	$7.65_{-0.37}^{+0.36}$ & 	$6.15_{-0.89}^{+1.04}$	&10.21	\\
\smallskip
6203980116	&16.00	&$1.398_{-0.021}^{+0.022}$ &	$0.212_{-0.044}^{+0.055}$ &	$7.47_{-0.56}^{+0.52}$ & 	$6.59_{-1.48}^{+1.81}$	&6.66\\	
\smallskip
6703010106	&19.15	&$2.298_{-0.022}^{+0.023}$ &	$0.310_{-0.061}^{+0.077}$ &	$6.35_{-0.36}^{+0.33}$ & 	$7.42_{-1.53}^{+1.93}$	&8.90\\	
\smallskip
6703010106	&19.28	&$2.770_{-0.019}^{+0.019}$ &	$0.271_{-0.049}^{+0.065}$ &	$5.42_{-0.29}^{+0.28}$ & 	$10.23_{-1.91}^{+2.52}$	&9.28	\\
\smallskip
6203980118	&19.54	&$2.371_{-0.022}^{+0.022}$ &	$0.249_{-0.052}^{+0.070}$ &	$5.85_{-0.43}^{+0.38}$ & 	$9.53_{-2.09}^{+2.78}$	&6.87\\	
\smallskip
6203980118	&19.92	&$3.051_{-0.023}^{+0.023}$ &	$0.349_{-0.060}^{+0.080}$ &	$5.47_{-0.29}^{+0.28}$ & 	$8.74_{-1.56}^{+2.06}$	&9.29\\	
\smallskip
6203980118	&19.99	&$3.271_{-0.020}^{+0.020}$ &	$0.342_{-0.045}^{+0.053}$ &	$5.08_{-0.21}^{+0.21}$ & 	$9.55_{-1.31}^{+1.54}$	&11.84	\\
\smallskip
6203980119	&20.18	&$2.683_{-0.021}^{+0.021}$ &	$0.331_{-0.047}^{+0.059}$ &	$5.75_{-0.28}^{+0.27}$ & 	$8.11_{-1.21}^{+1.52}$	&10.32\\	
\smallskip
6203980119	&20.25	&$2.844_{-0.021}^{+0.023}$ &	$0.317_{-0.051}^{+0.061}$ &	$5.60_{-0.28}^{+0.26}$ & 	$8.98_{-1.53}^{+1.80}$	&10.16\\	
\smallskip
6203980119	&20.31	&$2.700_{-0.017}^{+0.017}$ &	$0.270_{-0.041}^{+0.050}$ &	$5.58_{-0.30}^{+0.27}$ & 	$10.00_{-1.60}^{+1.93}$	&9.32	\\
\smallskip
6203980119	&20.38	&$2.601_{-0.023}^{+0.024}$ &	$0.387_{-0.060}^{+0.066}$ &	$5.82_{-0.25}^{+0.25}$ & 	$6.73_{-1.10}^{+1.20}$	&11.41\\	
\smallskip
6203980119	&20.44	&$2.605_{-0.017}^{+0.018}$ &	$0.325_{-0.041}^{+0.045}$ &	$5.79_{-0.23}^{+0.22}$ & 	$8.01_{-1.05}^{+1.16}$	&12.78\\	
\smallskip
6203980119	&20.50	&$2.634_{-0.017}^{+0.017}$ &	$0.363_{-0.043}^{+0.052}$ &	$5.94_{-0.23}^{+0.23}$ & 	$7.26_{-0.91}^{+1.08}$	&12.65	\\
\smallskip
6203980119	&20.89	&$2.111_{-0.016}^{+0.016}$ &	$0.270_{-0.041}^{+0.047}$ &	$6.13_{-0.28}^{+0.27}$ & 	$7.82_{-1.23}^{+1.42}$	&11.08\\	
\smallskip
6203980119	&20.96	&$2.023_{-0.014}^{+0.014}$ &	$0.196_{-0.028}^{+0.035}$ &	$6.37_{-0.33}^{+0.31}$ & 	$10.32_{-1.55}^{+1.90}$	&9.75\\	
\smallskip
6203980120	&21.02	&$1.991_{-0.007}^{+0.007}$ &	$0.312_{-0.014}^{+0.015}$ &	$6.49_{-0.10}^{+0.10}$ & 	$6.39_{-0.31}^{+0.32}$	&32.19	\\
\hline\noalign{\smallskip}
\end{tabular} 
\end{center}
\label{Tab:qpo_ni}
Notes:\\ 
rms: root mean square; $\nu_0$: centroid frequency; $\Delta$: half width at half maximum; $\sigma$: significance; QPO: quasiperiodic oscillation\\
$^{\dagger}$: since 60180 MJD
\end{table*}

\addtocounter{table}{-1}
\begin{table*}
\caption{continued}
\begin{center}
\begin{tabular}{rrrrrrr}
\hline\noalign{\smallskip}
\multicolumn{1}{c}{Obs ID} & \multicolumn{1}{c}{day$^{\dagger}$} & \multicolumn{1}{c}{$\nu_{\mr{0; QPO1}}$ (Hz)} & \multicolumn{1}{c}{$\Delta_{\mr{QPO1}}$ (Hz)} & \multicolumn{1}{c}{rms$_{\mr{QPO1}}$ (\%)}   &  \multicolumn{1}{c}{Q$_{\mr{QPO1}}$}  & \multicolumn{1}{c}{$\sigma_{\mr{QPO1}}$}  \\ 
\hline\noalign{\smallskip}
6203980121	&22.05	&$2.337_{-0.011}^{+0.012}$ &	$0.313_{-0.032}^{+0.034}$ &	$5.91_{-0.16}^{+0.16}$ & 	$7.46_{-0.79}^{+0.85}$	&18.09\\	
\smallskip
6703010107	&22.27	&$2.557_{-0.011}^{+0.011}$ &	$0.470_{-0.024}^{+0.025}$ &	$5.92_{-0.10}^{+0.10}$ & 	$5.45_{-0.30}^{+0.31}$	&29.32\\	
\smallskip
6203980122	&23.17	&$2.548_{-0.022}^{+0.022}$ &	$0.267_{-0.043}^{+0.054}$ &	$5.73_{-0.34}^{+0.32}$ & 	$9.55_{-1.62}^{+2.01}$	&8.43	\\
\smallskip
6203980122	&23.48	&$3.154_{-0.026}^{+0.028}$ &	$0.424_{-0.069}^{+0.085}$ &	$4.95_{-0.26}^{+0.25}$ & 	$7.45_{-1.28}^{+1.56}$	&9.59\\	
\smallskip
6203980130	&39.44	&$6.173_{-0.058}^{+0.030}$ &	$1.512_{-0.148}^{+0.158}$ &	$2.03_{-0.07}^{+0.07}$ & 	$4.08_{-0.44}^{+0.45}$	&13.83\\	
\smallskip
6203980131	&40.02	&$6.826_{-0.064}^{+0.065}$ &	$1.282_{-0.207}^{+0.236}$ &	$1.63_{-0.08}^{+0.08}$ & 	$5.33_{-0.91}^{+1.03}$	&9.66	\\
\smallskip
6703010113	&40.28	&$6.831_{-0.046}^{+0.046}$ &	$1.581_{-0.111}^{+0.120}$ &	$1.62_{-0.04}^{+0.03}$ & 	$4.32_{-0.33}^{+0.34}$	&22.85\\	
\smallskip
6703010114	&41.12	&$9.323_{-0.657}^{+0.321}$ &	$0.032_{-0.032}^{+3.599}$ &	$0.59_{-0.36}^{+0.34}$ & 	$295.42_{-316.25}^{+33700.73}$	&0.83\\	
\smallskip
6557020401	&41.19	&$6.849_{-0.033}^{+0.033}$ &	$1.513_{-0.105}^{+0.099}$ &	$1.53_{-0.03}^{+0.03}$ & 	$4.53_{-0.34}^{+0.32}$	&22.33	\\
\smallskip
6557020402	&42.02	&$7.007_{-0.093}^{+0.089}$ &	$1.150_{-0.290}^{+0.310}$ &	$1.47_{-0.11}^{+0.10}$ & 	$6.09_{-1.62}^{+1.72}$	&6.67\\	
\smallskip
6703010115	&42.22	&$7.057_{-0.104}^{+0.571}$ &	$0.325_{-0.308}^{+0.915}$ &	$0.41_{-0.11}^{+0.12}$ & 	$21.68_{-20.83}^{+62.73}$	&1.83\\	
\hline\noalign{\smallskip}
\end{tabular} 
\end{center}
Notes:\\ 
rms: root mean square; $\nu_0$: centroid frequency; $\Delta$: half width at half maximum; $\sigma$: significance; QPO: quasiperiodic oscillation\\
$^{\dagger}$: since 60180 MJD
\end{table*}

\begin{table*}
\caption{Parameters of the harmonic QPO in the PDS derived from \nice\ data}
\begin{center}
\begin{tabular}{rrrrrrr}
\hline\noalign{\smallskip}
\multicolumn{1}{c}{Obs ID} & \multicolumn{1}{c}{day$^{\dagger}$} & \multicolumn{1}{c}{$\nu_{\mr{0; QPO2}}$ (Hz)} & \multicolumn{1}{c}{$\Delta_{\mr{QPO2}}$ (Hz)} & \multicolumn{1}{c}{rms$_{\mr{QPO2}}$ (\%)}   &  \multicolumn{1}{c}{Q$_{\mr{QPO2}}$}  & \multicolumn{1}{c}{$\sigma_{\mr{QPO2}}$}  \\ 
\hline\noalign{\smallskip}
6203980104	&4.06	&$1.088_{-1.088}^{+0.360}$ &	$1.776_{-0.692}^{+0.274}$ &	$6.88_{-2.69}^{+1.49}$ & 	$0.61_{-0.85}^{+0.30}$	&1.28		\\	
\smallskip
6203980105	&5.04	&$1.423_{-0.234}^{+0.173}$ &	$1.984_{-0.256}^{+0.216}$ &	$6.01_{-0.77}^{+1.09}$ & 	$0.72_{-0.21}^{+0.16}$	&3.89		\\	
\smallskip
6703010101	&5.55	&$1.720_{-0.047}^{+0.037}$ &	$0.641_{-0.168}^{+0.270}$ &	$3.99_{-0.39}^{+0.53}$ & 	$2.68_{-0.77}^{+1.19}$	&5.16		\\	
\smallskip
6203980106	&6.01	&$1.878_{-0.052}^{+0.051}$ &	$2.263_{-0.055}^{+0.055}$ &	$5.98_{-0.13}^{+0.13}$ & 	$0.83\pm0.04$	&22.63		\\	
\smallskip
6203980107	&7.54	&$1.948_{-0.065}^{+0.065}$ &	$2.284_{-0.080}^{+0.076}$ &	$5.86_{-0.18}^{+0.17}$ & 	$0.85\pm0.06$	&15.91		\\	
\smallskip
6203980108	&8.00	&$2.163_{-0.043}^{+0.043}$ &	$2.308_{-0.054}^{+0.053}$ &	$5.57_{-0.10}^{+0.10}$ & 	$0.94\pm0.04$	&27.59		\\	
\smallskip
6203980109	&9.03	&$2.210_{-0.037}^{+0.037}$ &	$2.369_{-0.045}^{+0.045}$ &	$5.54_{-0.08}^{+0.08}$ & 	$0.93\pm0.03$	&33.61		\\	
\smallskip
6203980110	&10.52	&$2.320_{-0.048}^{+0.046}$ &	$2.335_{-0.061}^{+0.062}$ &	$5.45_{-0.10}^{+0.11}$ & 	$0.99\pm0.05$	&26.31		\\	
\smallskip
6203980111	&11.04	&$1.995_{-0.046}^{+0.045}$ &	$2.413_{-0.051}^{+0.051}$ &	$5.89_{-0.11}^{+0.11}$ & 	$0.83\pm0.04$	&26.83		\\	
\smallskip
6703010103	&11.35	&$1.894_{-0.095}^{+0.089}$ &	$2.278_{-0.105}^{+0.101}$ &	$6.02_{-0.26}^{+0.26}$ & 	$0.83\pm0.08$	&11.53		\\	
\smallskip
6203980112	&12.06	&$1.677_{-0.168}^{+0.158}$ &	$2.189_{-0.213}^{+0.171}$ &	$6.25_{-0.60}^{+0.54}$ & 	$0.77_{-0.15}^{+0.13}$	&5.23		\\	
\smallskip
6203980112	&12.20	&$1.095_{-1.095}^{+0.113}$ &	$2.178_{-0.175}^{+0.192}$ &	$8.10_{-0.39}^{+0.40}$ & 	$0.50_{-0.54}^{+0.10}$	&10.40		\\	
\smallskip
6203980112	&12.33	&$2.249_{-0.196}^{+0.177}$ &	$2.241_{-0.236}^{+0.229}$ &	$5.57_{-0.44}^{+0.45}$ & 	$1.00_{-0.19}^{+0.18}$	&6.27		\\	
\smallskip
6203980112	&12.46	&$2.596_{-0.147}^{+0.133}$ &	$2.342_{-0.232}^{+0.225}$ &	$5.29_{-0.29}^{+0.27}$ & 	$1.11_{-0.17}^{+0.16}$	&9.17		\\	
\smallskip
6203980112	&12.58	&$2.838_{-0.156}^{+0.144}$ &	$2.158_{-0.277}^{+0.267}$ &	$4.91_{-0.34}^{+0.31}$ & 	$1.32_{-0.24}^{+0.23}$	&7.32		\\	
\smallskip
6203980112	&12.65	&$2.548_{-0.171}^{+0.151}$ &	$2.297_{-0.218}^{+0.224}$ &	$5.28_{-0.31}^{+0.33}$ & 	$1.11_{-0.18}^{+0.17}$	&8.48		\\	
\smallskip
6203980112	&12.71	&$2.627_{-0.176}^{+0.166}$ &	$2.633_{-0.236}^{+0.238}$ &	$5.09_{-0.30}^{+0.29}$ & 	$1.00_{-0.16}^{+0.15}$	&8.45		\\	
\smallskip
6203980112	&12.77	&$2.434_{-0.173}^{+0.155}$ &	$2.333_{-0.210}^{+0.208}$ &	$5.50_{-0.34}^{+0.33}$ & 	$1.04_{-0.17}^{+10.16}$	&8.17		\\	
\smallskip
6203980112	&12.84	&$2.465_{-0.127}^{+0.118}$ &	$2.640_{-0.141}^{+0.144}$ &	$5.39_{-0.21}^{+0.22}$ & 	$0.93\pm0.10$	&12.77		\\	
\smallskip
6203980112	&12.90	&$2.610_{-0.110}^{+0.102}$ &	$2.401_{-0.141}^{+0.144}$ &	$5.36_{-0.19}^{+0.20}$ & 	$1.09\pm0.11$	&13.80		\\	
\smallskip
6203980113	&13.42	&$2.520_{-0.053}^{+0.053}$ &	$2.440_{-0.069}^{+0.067}$ &	$5.29_{-0.11}^{+0.10}$ & 	$1.03\pm0.05$	&24.82		\\	
\smallskip
6203980114	&14.00	&$2.571_{-0.051}^{+0.050}$ &	$2.545_{-0.059}^{+0.060}$ &	$5.28_{-0.09}^{+0.09}$ & 	$1.01\pm0.04$	&28.56		\\	
\smallskip
6703010104	&14.71	&$2.859_{-0.122}^{+0.117}$ &	$2.349_{-0.165}^{+0.168}$ &	$4.98_{-0.22}^{+0.22}$ & 	$1.22\pm0.14$	&11.18		\\	
\smallskip
6203980115	&15.61	&$2.363_{-0.130}^{+0.121}$ &	$2.459_{-0.152}^{+0.153}$ &	$5.68_{-0.25}^{+0.26}$ & 	$0.96\pm0.11$	&11.24		\\	
\smallskip
6203980116	&16.00	&$2.485_{-0.212}^{+0.193}$ &	$2.184_{-0.452}^{+0.320}$ &	$5.03_{-0.64}^{+0.42}$ & 	$1.14_{-0.33}^{+0.25}$	&3.92		\\	
\smallskip
6703010106	&19.15	&$4.633_{-0.181}^{+0.167}$ &	$2.096_{-0.414}^{+0.469}$ &	$3.20_{-0.27}^{+0.26}$ & 	$2.21_{-0.52}^{+0.57}$	&5.84		\\	
\smallskip
6703010106	&19.28	&$5.400_{-0.224}^{+0.173}$ &	$2.040_{-0.484}^{+0.611}$ &	$2.99_{-0.25}^{+0.25}$ & 	$2.65_{-0.74}^{+0.88}$	&5.93		\\	
\smallskip
6203980118	&19.54	&$4.484_{-0.229}^{+0.175}$ &	$1.445_{-0.457}^{+0.616}$ &	$2.89_{-0.37}^{+0.36}$ & 	$3.10_{-1.14}^{+1.44}$	&3.88		\\	
\smallskip
6203980118	&19.92	&$6.043_{-0.137}^{+0.127}$ &	$1.490_{-0.454}^{+0.605}$ &	$2.51_{-0.23}^{+0.23}$ & 	$4.06_{-1.33}^{+1.73}$	&5.48		\\	
\smallskip
6203980118	&19.99	&$6.451_{-0.089}^{+0.093}$ &	$0.903_{-0.243}^{+0.309}$ &	$2.06_{-0.17}^{+0.16}$ & 	$7.15_{-2.02}^{+2.55}$	&6.12		\\	
\smallskip
6203980119	&20.18	&$5.139_{-0.172}^{+0.140}$ &	$2.172_{-0.365}^{+0.427}$ &	$3.17_{-0.18}^{+0.18}$ & 	$2.37_{-0.48}^{+0.53}$	&8.75		\\	
\smallskip
6203980119	&20.25	&$5.415_{-0.153}^{+0.139}$ &	$1.806_{-0.339}^{+0.409}$ &	$2.97_{-0.21}^{+0.20}$ & 	$3.00_{-0.65}^{+0.76}$	&7.13		\\	
\smallskip
6203980119	&20.31	&$5.381_{-0.091}^{+0.092}$ &	$1.014_{-0.249}^{+0.293}$ &	$2.47_{-0.23}^{+0.21}$ & 	$5.31_{-1.39}^{+1.62}$	&5.38		\\	
\smallskip
6203980119	&20.38	&$5.080_{-0.146}^{+0.127}$ &	$2.058_{-0.307}^{+0.351}$ &	$3.30_{-0.18}^{+0.18}$ & 	$2.47_{-0.44}^{+0.48}$	&9.14		\\	
\smallskip
6203980119	&20.44	&$5.091_{-0.107}^{+0.087}$ &	$1.745_{-0.313}^{+0.388}$ &	$2.92_{-0.15}^{+0.15}$ & 	$2.92_{-0.58}^{+0.70}$	&9.82		\\	
\smallskip
6203980119	&20.50	&$5.247_{-0.083}^{+0.078}$ &	$1.852_{-0.253}^{+0.296}$ &	$3.04_{-0.13}^{+0.13}$ & 	$2.83_{-0.43}^{+0.49}$	&11.86		\\	
\smallskip
6203980119	&20.89	&$4.103_{-0.143}^{+0.118}$ &	$2.112_{-0.326}^{+0.346}$ &	$3.66_{-0.23}^{+0.23}$ & 	$1.94\pm0.37$	&7.89		\\	
\smallskip
6203980119	&20.96	&$3.701_{-0.162}^{+0.137}$ &	$2.655_{-0.309}^{+0.316}$ &	$4.36_{-0.22}^{+0.22}$ & 	$1.39\pm0.22$	&10.07		\\	
\smallskip
6203980120	&21.02	&$3.827_{-0.040}^{+0.039}$ &	$2.271_{-0.099}^{+0.100}$ &	$4.13_{-0.07}^{+0.07}$ & 	$1.69\pm0.09$	&28.04		\\	
\smallskip
6203980121	&22.05	&$4.465_{-0.082}^{+0.072}$ &	$2.063_{-0.208}^{+0.228}$ &	$3.60_{-0.12}^{+0.12}$ & 	$2.16_{-0.26}^{+0.27}$	&14.80		\\	
\hline\noalign{\smallskip}
\end{tabular} 
\end{center}
\label{Tab:qpo_ni_uh}
Notes:\\ 
rms: root mean square; $\nu_0$: centroid frequency; $\Delta$: half width at half maximum; $\sigma$: significance; QPO: quasiperiodic oscillation\\
$^{\dagger}$: since 60180 MJD
\end{table*}

\addtocounter{table}{-1}
\begin{table*}
\caption{continued}
\begin{center}
\begin{tabular}{rrrrrrr}
\hline\noalign{\smallskip}
\multicolumn{1}{c}{Obs ID} & \multicolumn{1}{c}{day$^{\dagger}$} & \multicolumn{1}{c}{$\nu_{\mr{0; QPO2}}$ (Hz)} & \multicolumn{1}{c}{$\Delta_{\mr{QPO2}}$ (Hz)} & \multicolumn{1}{c}{rms$_{\mr{QPO2}}$ (\%)}   &  \multicolumn{1}{c}{Q$_{\mr{QPO2}}$}  & \multicolumn{1}{c}{$\sigma_{\mr{QPO2}}$}  \\ 
\hline\noalign{\smallskip}
6703010107	&22.27	&$4.970_{-0.047}^{+0.046}$ &	$2.427_{-0.112}^{+0.116}$ &	$3.42_{-0.06}^{+0.06}$ & 	$2.05_{-0.11}^{+0.12}$	&28.76		\\	
\smallskip
6203980122	&23.17	&$4.995_{-0.165}^{+0.127}$ &	$2.370_{-0.516}^{+0.578}$ &	$3.45_{-0.20}^{+0.20}$ & 	$2.11_{-0.53}^{+0.56}$	&8.81		\\	
\smallskip
6203980122	&23.48	&$6.280_{-0.111}^{+0.115}$ &	$1.245_{-0.286}^{+0.334}$ &	$2.31_{-0.18}^{+0.17}$ & 	$5.05_{-1.25}^{+1.45}$	&6.56		\\	
\smallskip
6203980130	&39.44	&$13.753_{-1.787}^{+0.642}$ &	$0.049_{-0.045}^{+2.449}$ &	$0.38_{-0.13}^{+0.14}$ & 	$\ge5.62$	&1.43\\			
\smallskip
6203980131	&40.02	&$12.489_{-0.272}^{+1.349}$ &	$0.013_{-0.013}^{+2.297}$ &	$0.33_{-0.17}^{+0.11}$ & 	$\ge5.44$	&0.96\\			
\smallskip
6703010113	&40.28	&$12.708_{-0.122}^{+0.548}$ &	$\le0.738$ &	$0.23_{-0.08}^{+0.06}$ & 	$\ge17.21$	&1.44		\\	
\smallskip
6703010114	&41.12	&$4.310_{-1.031}^{+1.359}$ &	$3.713_{-2.526}^{+2.845}$ &	$1.32_{-0.40}^{+0.27}$ & 	$1.16_{-1.07}^{+1.26}$	&1.64		\\	
\smallskip
6557020401	&41.19	&$13.748_{-1.619}^{+0.004}$ &	$0.001_{-0.001}^{+0.756}$ &	$0.25_{-0.07}^{+0.06}$ & 	$\ge18.20$&	1.69\\	
\smallskip
6557020402	&42.02	&$15.140_{-1.403}^{+0.489}$ &	$0.009_{-0.008}^{+2.922}$ &	$0.46_{-0.17}^{+0.18}$ & 	$\ge5.18$	&1.33\\			
\smallskip
6703010115	&42.22	&$2.753_{-0.627}^{+0.366}$ &	$0.752_{-0.702}^{+1.241}$ &	$0.37_{-0.17}^{+0.14}$ & 	$3.66_{-4.25}^{+6.52}$	&1.09		\\	
\hline\noalign{\smallskip}
\end{tabular} 
\end{center}
Notes:\\ 
rms: root mean square; $\nu_0$: centroid frequency; $\Delta$: half width at half maximum; $\sigma$: significance; QPO: quasiperiodic oscillation\\
$^{\dagger}$: since 60180 MJD
\end{table*}

\end{appendix}

\end{document}